\documentclass[aps,pre,floats,twocolumn,showpacs]{revtex4}

\usepackage{epsfig}

\begin{document}

\title{Topology of the World Trade Web}

\author{M$^{\mbox{\underline{a}}}$ \'Angeles Serrano and Mari{\'a}n Bogu{\~n}{\'a}}

\affiliation{Departament de F{\'\i}sica Fonamental, Universitat de
  Barcelona, Av. Diagonal 647, 08028 Barcelona, Spain}

\date{\today}

\begin{abstract}
Economy, and consequently trade, is a fundamental part of human
social organization which, until now, has not been studied within
the network modelling framework. Networks are mathematical tools
used in the modelling of a wide variety of systems in social and
natural science. Examples of these networks range from metabolic
and cell networks to technological webs. Here we present the first
empirical characterization of the world trade web, that is, the
network built upon the trade relationships between different
countries in the world. This network displays the typical
properties of complex networks, namely, scale-free degree
distribution, the {\it small world} property, a high clustering
coefficient and, in addition, degree-degree correlation between
different vertices. All these properties make the world trade web
a complex network, which is far from being well-described through
a classical random network description.
\end{abstract}


\maketitle

The world is facing a challenging era. Social, political and
economic arrangements initiated after the end of the Second World
War are now culminating in the recognition of globalization, a
process which has been accelerated by the new technological
advances. When applied to the international economic order,
globalization involves control of capital flow and liberalization
of trade. As a consequence, economies around the world are
becoming more and more interrelated, in other words, the world is
becoming a {\it global-village} \cite{krugman,amin}. In this
scenario, trade plays a central role as one of the most important
interaction channels between countries. The relevance of the
international trade system goes beyond the fundamental exchange of
goods and services. For instance, it can also be the channel for
crises spreading \cite{glick}. A good example is found in the
recent Asiatic crisis, which shows how economic perturbations
originated in a country can somehow propagate elsewhere in the
world \cite{noland,goldstein}. Thus, it seems natural to analyze
the world trade system at a global level, every country being
important regardless of its size or wealth. Despite the extremely
complex nature of the problem, relevant structural information can
be extracted from modelling the system as a network, where
countries are represented as vertices and trade channels as links
between these vertices. In this way, the global trade system can
be examined under a topological point of view. This analysis will
reveal complex properties which cannot be explained by the
classical random graph theory.

Complex networks have been the subject of an intense research
activity over the last years \cite{barabasi01,dorogorev}. This
great interest is fully justified by the extremely important role
that this class of systems play in many different fields. Examples
range from metabolic networks, where cell functionality is
sustained by the network structure, to technological webs, where
topology determines the system ability transmitting information
\cite{jeong00,barab02,alexei2,albert99}. The term complex network
typically refers to networks showing the following properties:
(i) scale-free (SF) degree distribution, $P(k)\sim k^{-\gamma}$
with $2<\gamma\leq 3$, where the degree, $k$, is defined as the
number of edges emanating from a vertex,
(ii) the {\it small-world} property \cite{watts98}, which states
that the average path length between any pair of vertices grows
logarithmically with the system size and
(iii) a high clustering coefficient, that is, the neighbors of a
given vertex are interconnected with high probability.
In addition, degree-degree correlation has been recently added to
this list since it appears as a common feature in many real-world
networked systems \cite{assortative,alexei2,romu,goh01b}. This
correlation accounts for the probability that a vertex of degree
$k$ is connected to a vertex of degree $k'$ and is a key issue for
the correct description of the hierarchical organization within
the network. This correlation is found to be assortative, that is,
highly connected vertices tend to attach to other highly connected
vertices in social networks such as scientific collaboration
networks \cite{assortative}; conversely, the correlation is
disassortative in technological networks, such as the Internet
\cite{alexei2}.

Classical random graph theory, first studied by Erd{\"o}s and
R{\'e}nyi \cite{erdos59,erdos60}, does not provide a good
framework to fit all the above properties. This fact has posed the
question of the origin of these anomalous topological features.
Two possible mechanisms could explain their appearance: either the
network is the result of macroscopic constraints, that is, the
network is made {\it ad hoc} so that those properties are
satisfied, or there is a self-organized evolution process leading,
in the stationary state, to complex structures. This idea is at
the core of the preferential attachment mechanisms, first
introduced by Barab\'asi and Albert \cite{barab99}, where a
dynamical process of creation of new links, or rewiring of the
existing ones, using global (or quasi global) information leads to
SF networks displaying complex properties.

Among all the studied networks, the social and technological
networks have become the paradigmatic example of complex networks.
Perhaps, the reason lies in the fact that each type exemplifies
human activity working at one of two different levels: the
cooperative level, where concepts such as friendship are dominant,
and the competitive level, where the activity is governed by
optimization criteria. These two different levels will probably
lead to different attachment mechanisms which could be the origin
of the assortative/disassortative degree-degree correlation
observed in these networks. As an example of social network
working at the competitive level, we study the network of trade
relationships between different countries in the world, hereafter
referred to as the world trade web (WTW). The topological
characterization of the WTW is of primary interest for the
modelling of crisis propagation at the global level as well as for
the understanding of the effects that the new liberalist policies
have on the world trade system. Moreover, we shall show that the
WTW is a complex network sharing many properties with
technological networks.

\begin{figure}[t]
   \epsfig{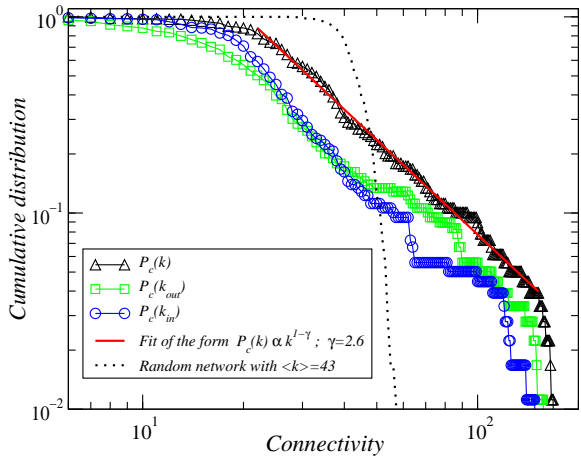}
     \caption{Cumulative in- and out-degree distributions,
     $P_c(k_{in})$ and $P_c(k_{out})$, and undirected, $P_c(k)$, corresponding to
     the import/export world trade web. The solid line is a power law fit of
     the form $P_c(k)\sim k^{\gamma-1}$ with $\gamma=2.6\pm0.1$. It is also shown
     the cumulative distribution of the equivalent random network with the same
     average degree}
  \label{survival}
\end{figure}

In order to perform our analysis, we extracted data from aggregated trade
statistics tables in the International Trade Center site \cite{ITC}, which are
based on the COMTRADE database of the United Nations Statistics Division. These
tables contain, for each country, an import and an export list detailing the
forty more important exchanged merchandises in the year 2000. Primary and
secondary markets are also reported for each product. If we consider imports as
in-degrees and exports as out-degrees, it is possible to construct a directed
network where vertices represent countries and directed links represent the
import/export relations between them. The fact that the number of merchandises
is bounded is, {\it a priori}, a limitation for the analysis. However, it is
possible to overcome this problem taking advantage of the symmetry between in
and out degrees. Let $\tilde{A}^{imp}_{ij}$ and $\tilde{A}^{exp}_{ij}$ be the
import/export adjacency matrices calculated from the import/export databases.
Each adjacency matrix is defined so that $\tilde{A}_{ij}=1$ if the country $i$
imports from/exports to the country $j$ and zero otherwise. These matrices
account only for a subset out of the total number of actual connections between
countries. The import/export connections that are of little relevance for a
given country are not considered in the matrices $\tilde{A}$, although they may
be relevant to the partners as the symmetric export/import links. In fact,
imports and exports definitions can apply to the same trade flow depending on
whether origin or destination is considered. This implies that the complete
adjacency matrices satisfy the symmetry relation $A^{imp}_{ij}=A^{exp}_{ji}$,
which can be used in order to recover missing information from the original
matrices. Thus, we can write
\begin{equation}
A^{imp}_{ij}=\frac{1}
{1+\delta_{(\tilde{A}^{imp}_{ij}+\tilde{A}^{exp}_{ji}),2}}
\left[\tilde{A}^{imp}_{ij}+\tilde{A}^{exp}_{ji}\right]
\end{equation}
where $\delta_{\cdot,\cdot}$ is the Kronecker delta function. In
this way we obtain an adjacency matrix where each connection is
relevant, at least, to one of the two involved countries. At this
point, it is worth noticing that we consider the unweighed version
of the WTW. Since the weight of a link can be different depending
on the import/export point of view, it is  unclear how these
weights should be assigned. After this symmetrization procedure,
we obtain a directed network with 179 vertices representing
countries and 7510 directed links representing commercial channels
among them. The average degree of this network is $\langle k_{in}
\rangle = \langle k_{out} \rangle =30.9$.

\begin{figure}[t]
   \epsfig{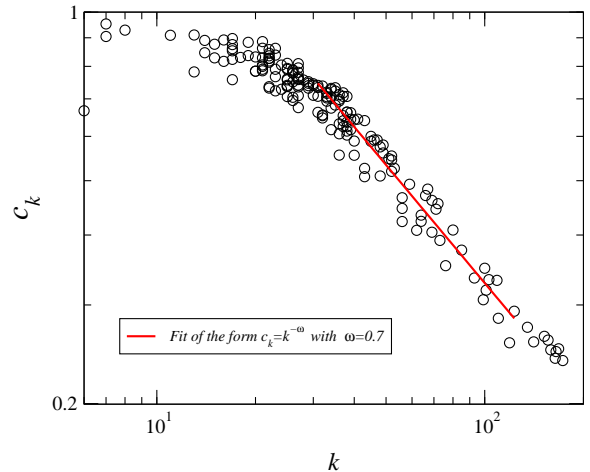}
     \caption{Clustering coefficient of single countries as a function of their
     degree for the undirected version of the WTW. The solid line is a power
     law fit of the form $c_k \sim k^{-\omega}$ with $\omega=0.7\pm0.05$}
  \label{clustering}
\end{figure}

The question that first arises refers to the directed nature of
the WTW. In fact, the in- and the out-degree of a given vertex are
random quantities which may be correlated. A complete description
should involve the knowledge of the joint probability
$p(k_{in},k_{out})$. This function is often difficult to obtain
although relevant information can be extracted from the
correlation coefficient $r=(\langle k_{in} k_{out} \rangle
-\langle k \rangle^2)/\sigma_{in} \sigma_{out}$, where
$\sigma_{in}$/$\sigma_{out}$ are the in/out standard deviations
(notice that $\langle k_{in} \rangle=\langle k_{out} \rangle
=\langle k \rangle$). For the WTW this coefficient is $r=0.91$,
pointing to a strong similarity in the number of in and out
connections. However, not all these connections run in both
directions, that is, the fact that country A imports from country
B does not necessary imply that country B imports from country A.
In order to quantify this effect we compute the reciprocity of the
network, defined as the fraction of links pointing simultaneously
to both ends of the link. In our case the reciprocity is 0.61.
These results suggest that, actually, the WTW may be thought of as
an undirected network without losing relevant topological
information. The average degree for the undirected version of the
WTW is $\langle k \rangle=43$.

One of the most important topological properties of a network is
the degree distribution, $P(k)$. This quantity measures the
probability of a randomly chosen vertex to have $k$ connections to
other vertices. In our case, due to the directed nature of the
network, we have to distinguish between in- and out-degree
distributions. Fig.\ref{survival} shows the in, out, and
undirected cumulative distributions for the WTW, defined as
$P_c(k) \equiv \sum_{k'=k} P(k')$. In all cases, the cumulative
distribution shows a flat approach to the origin, indicating the
presence of a maximum in $P(k)$, at $k \sim 20$, and, at this
respect, similar to the Erd{\"o}s-R{\'e}nyi network. However, for
$k>20$ the cumulative distribution is followed by a power law
decay $P_c(k) \sim k^{1-\gamma}$, with $\gamma \approx 2.6$,
showing a strong deviation from the exponential tail predicted by
the classical random graph theory. The exponent $\gamma$ is found
to be within the range defined by many other complex networks and,
thus, we can state that the WTW belongs to the recently identified
class of SF networks \cite{barabasi01,dorogorev}. The SF property
implies an extremely high level of degree heterogeneity. Indeed,
the second moment of the degree distribution, $\langle k^2
\rangle$, diverges in the thermodynamic limit for any SF network.

It may be surprising that, in fact, the SF region does not extend
to the whole degree domain and that, for small values of the
degree, the distribution is similar to the Erd{\"o}s-R{\'e}nyi
network. In fact, the underlying preferential attachment
mechanisms could differ depending on the particular political and
economic situation of a country. Low-degree countries, most of
which turn out to be the poorest, are basically constrained to
subsistence trade flows and, therefore, preferential attachment
mechanisms could not hold. As expected, there exists a positive
correlation between the number of trade channels of a country and
its wealth, measured by the {\it per capita} Gross Domestic
Product (GDP) \cite{gdp}. This correlation is found to be high,
0.65, which means that, indeed, most low-connected countries are
poor countries --Angola, Somalia, Rwanda, Cambodia...-- and most
high-connected countries are rich countries --the USA, Japan,
Germany and the UK, for example. However, there also exists a
significant number of cases in the reversal situation, that is,
low {\it per capita} GDP countries with a large number of
connections and high {\it per capita} GDP countries with a
relatively low number of trade channels. A germane example for the
first circumstance is that of Norway or Iceland, which are between
the top ten wealthier countries but only have 56 and 24 trade
channels respectively. For the second case Brazil, China or Russia
are typical examples.

Thanks to a number of recent studies, it is becoming more and more
evident that real networks are not completely random but they are
organized according to a hierarchical structure
\cite{alexei2,barab02,ravasz,goh01b}. This hierarchy is usually
analyzed by means of the local clustering coefficient and the
degree-degree correlation. The clustering coefficient of the
vertex $i$, of degree $k_i$, is defined as $c_i\equiv
2n_i/k_i(k_i-1)$, where $n_i$ is the number of neighbors of $i$
that are interconnected. If hierarchy was not present in the
system, the local clustering coefficient should be a random
quantity independent of any other property. Fig. \ref{clustering}
shows the local clustering coefficient of the undirected WTW as a
function of the vertex's degree. As it is clearly seen, this
function has a strong dependence on the vertex's degree, with a
power law behavior $c_k\sim k^{-\omega}$, with
$\omega=0.7\pm0.05$. The clustering coefficient averaged over the
whole network is $C=0.65$, greater by a factor 2.7 than the value
corresponding to a random network of the same size.

\begin{figure}[t]
   \epsfig{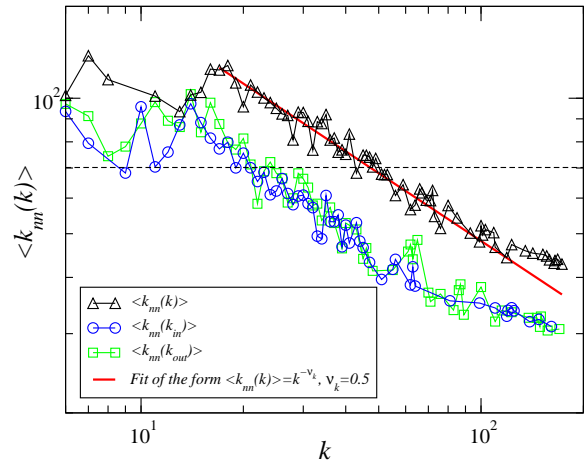}
     \caption{Average in (out, undirected) nearest neighbors degree as a function
     of the in (out, undirected) degree of the vertex. The solid line is a fit of
     the form $\langle k_{nn}(k)\rangle \sim k^{-\nu_{k}}$ with $\nu_{k}=0.5\pm0.05$.
     The dashed line is the theoretical value of the $\langle k_{nn} \rangle$ corresponding to
     an uncorrelated network, that is, $\langle k^2 \rangle /\langle k \rangle=70.18$.}
  \label{knn}
\end{figure}

Hierarchy is also reflected on the degree-degree correlation
through the conditional probability $P(k|k')$, which measures the
probability of a vertex of degree $k'$ to be linked to a vertex of
degree $k$. Again, this function is difficult to measure due to
statistical fluctuations. In order to characterize this
correlation, it is more useful to work with the average nearest
neighbors degree (ANND), defined as $\langle k_{nn} (k) \rangle
\equiv \sum_{k'} k'P(k'|k)$ \cite{romu}. For uncorrelated
networks, this function reads $\langle k_{nn} \rangle =\langle
k^2\rangle/\langle k\rangle$, independent of $k$. However, recent
studies have revealed that almost all real networks show
degree-degree correlation \cite{alexei2,assortative} which
translates into a $k$ dependence in the ANND. This correlation can
be assortative or disassortative depending on whether the ANND is
an increasing or decreasing function of the degree. Fig. \ref{knn}
reports the ANND for the directed and undirected versions of the
WTW. It can be observed a clear dependency on the vertex's degree,
with a power law decay $\langle k_{nn} (k) \rangle \sim
k^{-\nu_{k}}$ with $\nu_{k}=0.5\pm0.05$. This result means that
the WTW is a disassortative network where highly connected
vertices tends to connect to poorly connected vertices. This
result, together with the scaling law $c_k \sim k^{-\omega}$
reveals a hierarchical architecture of highly interconnected
countries that belong to influential areas which, in turn, connect
to other influential areas through hubs.

\begin{table}
  \centering
  \caption{}
  \begin{tabular}{|c|c|c|c|c|c|c|c|}
    \hline
     & size & $\langle k \rangle$ & $\langle d \rangle$ & $C$ & $\gamma$ & $\omega$ & $\nu_k$
     \\\hline \hline
    \text{WTW} & \text{ 179 } & \text{ 43 } & \text{ 1.8 } & \text{ 0.65 } & \text{ 2.6 } & \text{ 0.7 } & \text{ 0.5 } \\\hline
    \text{ Internet} & 5287 & 3.8 & 3.7 & 0.24 & 2.2 & 0.75 & 0.5 \\\hline
    \text{RG} & 179 & 43 & 1.73 & 0.24 & -- & 0 & 0 \\ \hline
  \end{tabular}
  \label{table1}
\end{table}

Surprisingly, these results point to a high similarity between the
WTW and the Internet. Indeed, the Internet is a SF network, with a
critical exponent $\gamma=2.2$, which is also organized in a
hierarchical fashion. The functional behavior found for the
clustering coefficient and the ANND is a power law decay as a
function of the degree with exponents $\omega_{int}=0.75$ and
$\nu_{k}=0.5$ \cite{alexei2}, exponents that turn out to be very
similar to the ones reported here for the WTW. In some sense,
these results are not surprising since both are competitive
systems evolving in a quasi free market and, in both cases, there
exists, for instance, a geographic limitation that increases the
connection costs and, thus, acts as a constraint in the
optimization process of each vertex. Table \ref{table1} presents a
summary of the main characteristics of the WTW, the Internet
\cite{alexei2} and a random graph of the same size and average
degree as the WTW.

As a final remark, the average path length, defined as the average
of the shortest distances between all the pairs of vertices, is
$\langle d \rangle=1.8$, which, in this case, is very similar to
the corresponding random network of the same size and average
degree.

In conclusion, this first approach to the topology of the WTW
points out some previously unnoticed features which are of primary
importance in the understanding of the new international order.
Our research suggests that the network's evolution is guided by
collective phenomena, and that self-organization plays a crucial
role in structuring the WTW scale-free inhomogeneities and its
hierarchical architecture. It remains an open question if these
properties could also be made apparent at other different scales,
for instance, in the trade relations between regions, cities or
even individuals. At the country level, the activity is driven by
competition. In the same way as it occurs with the Internet,
optimization criteria are applied to local decisions made by the
individual vertices. Resolutions are based on information that is
biased toward the {\it more visible} vertices, and may be
influenced by geographical or other convenience constraints. The
findings in this paper may lead to consider that there exist
underlying growing mechanisms common to all competitive systems,
characterized by disassortative associations, and such mechanisms
may differ from the evolutionary processes in social cooperative
networks, characterized by assortative associations.

Further modelling efforts must be done for acquiring a more
realistic representation of the WTW, where inward flows differ
from outward flows and where their weights depend on the exchanged
quantities. It is also essential to do further research on the
underlying formation mechanisms and on the dynamic processes
running on top of the WTW, such as economic crises spreading.

\begin{acknowledgments}
Acknowledgements are due to Romualdo Pastor-Satorras for helpful discussion and
advices. This work has been partially supported by the European commission FET
Open project COSIN IST-2001-33555.
\end{acknowledgments}

\end{document}